# Sub-kelvin cooling for the BICEP Array project

**L. Duband\*, T. Prouve\*, J. Bock\*\*,\*\*\*, L. Moncelsi\*\* and A. Schillaci\*\***

\*University Grenoble Alpes, CEA INAC-SBT F-38000 Grenoble, France
\*\* California Institute of Technology, Pasadena, CA 91125, USA
\*\*\* Jet Propulsion Laboratory, Pasadena, CA 91109, USA

**ABSTRACT**

In the field of astrophysics, the faint signal from distant galaxies and other dim cosmological sources at millimeter and submillimeter wavelengths require the use of high-sensitivity experiments. Cryogenics and the use of low-temperature detectors are essential to the accomplishment of the scientific objectives, allowing lower detector noise levels and improved instrument stability. Bolometric detectors are usually cooled to temperatures below 1K, and the constraints on the instrument are stringent, whether the experiment is a space-based platform or a ground-based telescope. The latter are usually deployed in remote and harsh environments such as the South Pole, where maintenance needs to be kept minimal.

CEA-SBT has acquired a strong heritage in the development of vibration-free multistage helium-sorption coolers, which can provide cooling down to 200mK when mounted on a cold stage at temperatures ≤ 5K. In this paper, we focus on the development of a three-stage cooler dedicated to the BICEP Array project led by Caltech/JPL, which aims to study the birth of the Universe and specifically the unique "B-mode" pattern imprinted by primordial gravitational waves on the polarization of the Cosmic Microwave Background. Several cryogenic receivers are being developed, each featuring one such helium-sorption cooler operated from a 4K stage cooled by a Cryomech pulse-tube with heat lifts of ≥1.35W at 4.2K and ≥36W at 45K. The major challenge of this project is the large masses to be cooled to sub-kelvin temperatures (26kg at 250mK) and the resulting long cooldown time, which in this novel cooler design is kept to a minimum with the implementation of passive and active thermal links between different temperature stages.

A first unit has been sized to provide 230, 70 and 2µW of net heat lifts at the maximum temperatures of 2.8K, 340 and 250mK, respectively, for a minimum duration of 48 hours. The unit has been manufactured, assembled and tested.

**INTRODUCTION**

The BICEP Array project is the next stage of the BICEP / Keck Array cosmic microwave background (CMB) polarization experiment. The scientific goal is to constrain the amplitude of a polarization signal with a characteristic 'B-mode' spatial pattern, associated with a primordial background of gravitational waves[1]. Current measurements from BICEP-Keck[2] are limited by experimental errors in carrying out a multi-component separation of CMB and Galactic signals. Thus, the next stage of the experiment must offer both significantly improved sensitivity, and wide frequency coverage to measure and subtract polarization Galactic foreground emission.

BICEP Array follows the BICEP architecture of using compact wide-field refractors (see Fig. 1) that have proven successful in controlling degree-scale systematic errors[3]. By increasing the telescope diameter from the 26 cm BICEP design to 55 cm, we substantially increase the optical étendue (AΩ product), by approximately a factor of 15 at 150 GHz. This permits 15 times more detectors at a given frequency with a corresponding improvement in sensitivity. BICEP Array will deploy four receivers in total to cover 6 observing frequencies at 30, 40, 95, 150, 220 and 270 GHz. The large étendue naturally results in a large 250 mK focal plane, with a diameter of 55 cm. Finally, the instrument is designed to shield the sensitive SQUID current amplifiers from magnetic fields using a combination of high-permeability and superconducting shields. For the focal plane, this includes a Nb shield mounted on the 350 mK stage, and individual Nb housings on the 2350 mK stage. The overall focal plane mass totals 54 kg, as given in Table 1.

The BICEP Array project involves a challenging cryogenic system. A Cryomech PT 415 is first used to cool several thermal shields and intercept heat loads at the 50 K and 4 K stages and provides the 4 K interface to the sorption cooler (see Fig. 1). The large heat load from the window is first dissipated by a multi-layer absorbing filter stack and an alumina filter at 50 K, and then further reduced by the lenses and additional filters at 4 K so that the absorbed optical power on the focal plane is small, nominally less than 1 µW. The focal plane is supported and thermally isolated on a multi-stage carbon-fiber truss structure with stages at 2 K, 0.35 K and 0.25 K. The multistage helium evaporative cooler cools the large focal plane masses at different temperature levels as presented in Table 1.

Helium evaporative sorption coolers can provide a wide range of heat lift capability at sub-kelvin temperature. They have no moving parts, are vibration less and can be designed to be self-contained and compact with a high duty cycle efficiency. Their robustness and the absence of any maintenance make them compatible with harsh environments. They rely on the capability of porous materials to adsorb or release a gas when cyclically cooled or heated. Using this physical process, one can design a compressor/pump which by managing the gas pressure in a closed system, can condense liquid at some appropriate location and then perform an evaporative pumping on the liquid bath to reduce its temperature. Consequently, it requires a pre-cooling stage at a temperature lower than the helium liquid-vapor transition. Several helium stages can be cascaded to allow operation from a 4-5 K heat sink. In this case the first helium 4 stage is used to provide cooling power at a temperature below 2 K and to condense the second stage filled with helium 3. This second stage can be in turn used as a pre-cooler for a third helium 3 stage allowing to reach temperature close to 200 mK.

**Table 1**: selected requirements for the sub-kelvin cooler

| Stage | Operating temperature | Net heat lift & hold time | Masses to be cooled |
|---|---|---|---|
| $^4$He stage<br>1$^{st}$ stage | 2.8 K or less | 230 µW for 48 h | 3 kg    aluminum<br>0.3 kg   stainless steel |
| Intercooler (IC)<br>$^3$He 2$^{nd}$ stage | 340 mK or less | 70 µW for 48 h | 15 kg   niobium<br>8 kg    copper<br>0.2 kg   stainless steel |
| Ultracooler (UC)<br>$^3$He 3$^{rd}$ stage | 250 mK or less | 2 µW at 250 mK<br>Must hold 48 h with 15 µW at ≈ 300 mK | 14 kg   copper<br>10 kg   niobium<br>2 kg    fiberglass<br>0.8 kg   stainless steel<br>0.7 kg   aluminum |

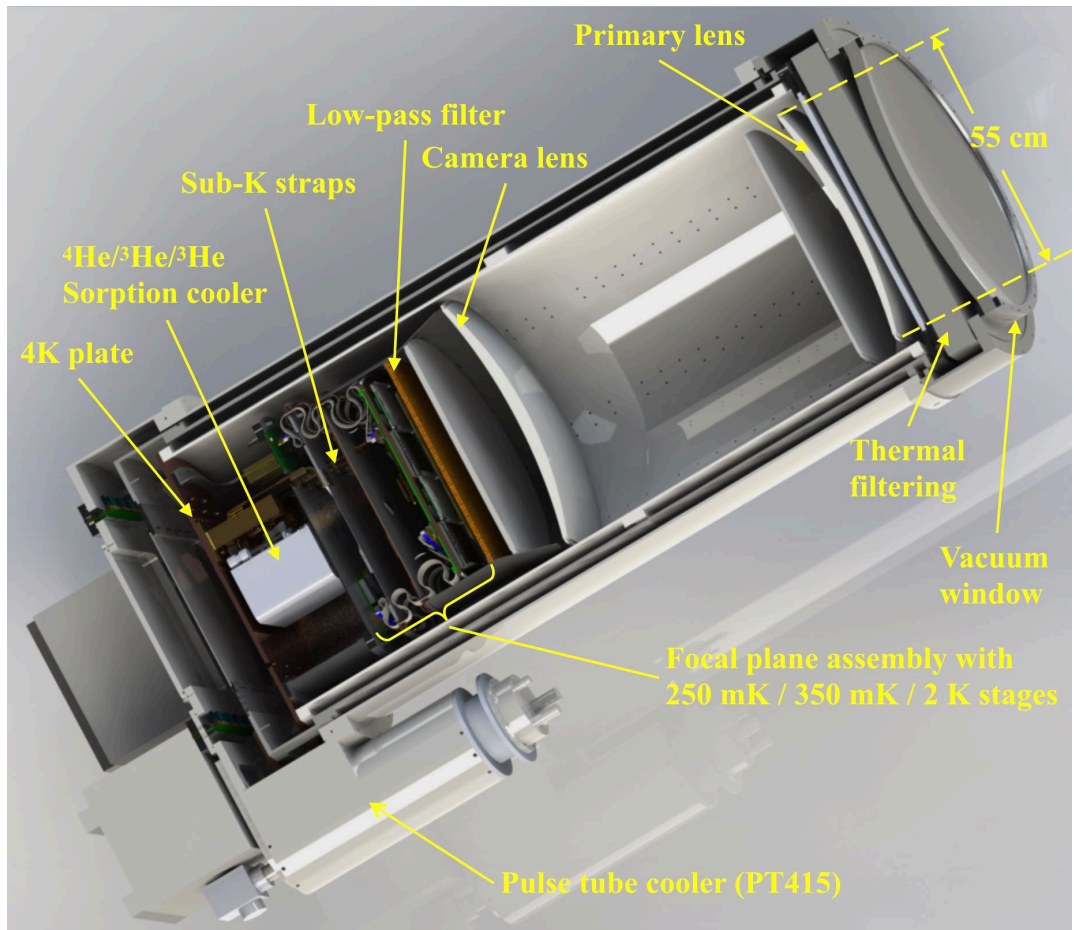

**Figure 1.** Schematic of the BICEP Array telescope

## COOLER DESCRIPTION

The cooler is a combination of 3 helium evaporative sorption coolers. This type of cooler has been described in numerous paper[4,5,6] and we only briefly recall here the basic principle (see Fig. 2). Two chambers, the evaporator and the charcoal sorption pump, are joined by thin-walled stainless steel tubes via the condenser, which is in good thermal contact with a cold heat sink at a temperature below the temperature of the critical point of helium. The cooler is charged with helium gas, and permanently sealed by a crimped tube. Once at the temperature of the heat sink, most of the helium is adsorbed by the activated charcoal in the sorption pump. The refrigeration cycle is initiated by heating the sorption pump by means of a resistive heater. This desorbs the gas from the charcoal, and once the gas pressure exceeds the saturated vapor pressure of helium at the temperature of the heat sink, liquid condenses in the condenser and falls into the evaporator. After condensation is complete, the sorption pump is activated by turning the heater off allowing the pump to cool through a passive thermal link or a heat switch to the heat sink temperature. The vapor pressure of helium is reduced and the temperature of the evaporator drops quickly to its operating temperature, where it remains stable until the liquid helium is exhausted. The refrigerator can be recycled indefinitely by heating the pump again.

For the BICEP Array project, the cooler is operated from a 4 K heat sink provided by a pulse tube mechanical cooler and comprises three stages. The helium 4 stage is first used to provide a lower temperature compatible with an efficient production of liquid helium 3 (typically below 2.5 K) to recycle the two $^3$He stages at the same time. This $^4$He stage is then used to provide cooling power at a temperature less than 2.8 K while the $^3$He stages are operating respectively below 340 and 250 mK. The UC cooler reaches very low temperature due to a reduction of the parasitic losses intercepted by the IC cooler. A general schematic is depicted on Fig. 3.

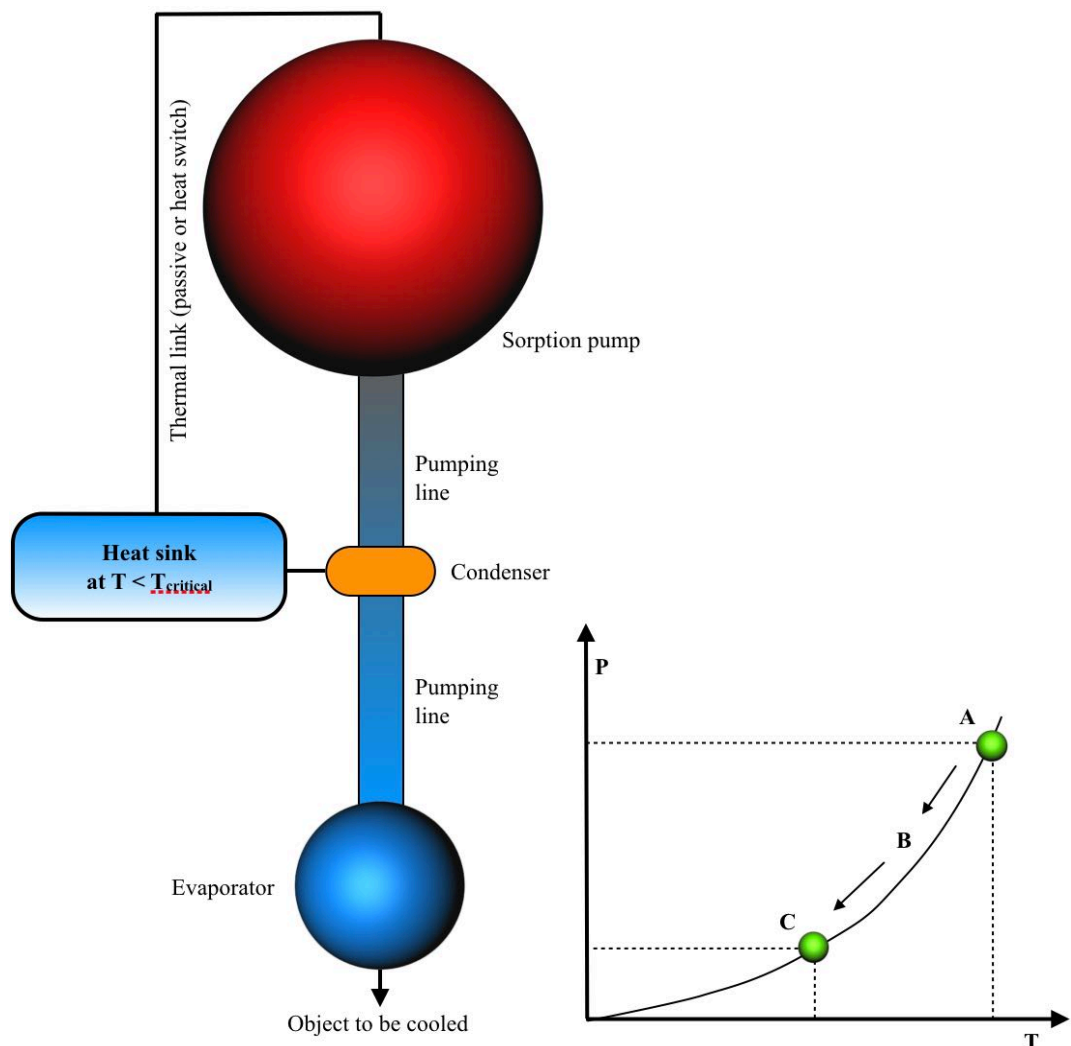

**Figure 2.** Principle of operation of a sorption cooler: [A] condensation, [B] cooldown and [C] low temperature

Different scenarios were evaluated for the recycling sequence, each one leading to a different sizing for the cooler. In order to keep the cooler compact and to limit the cost (amount of helium 3) a recycling phase featuring a double recycling was selected. In this case the $^4$He stage is recycled once to recycle the $^3$He stages, and then recycled again to provide cooling during 48 h. Thus, its operating temperature is much lower (below 2 K) and consequently the $^3$He stages are smaller because of lower parasitic loads. However, the recycling phase is slightly longer. Besides the required amount of helium needed to cope with the parasitic heat loads and net heat lifts for a hold time of 48 hours, the enthalpies of the various masses were also taken into account in the sizing.

The cool down time of the full experimental set-up from room temperature is an issue. Indeed, as indicated in table 1, each evaporator is connected to fairly massive items that can be made of aluminum, stainless steel, copper or niobium. By essence the evaporators are thermally isolated from each other and from the upper heat sinks; during the cool down process from ambient temperature, the thermal paths are mostly provided by convective effect due to the gas inside the tubes, pending it is possible to keep the sorption pumps hot enough to maintain the presence of gas as the temperature drops below 25 K. In that respect the heat switches on this cooler have been tuned so that their switching temperature is set in the 20 K range or above. Then several items were added to the design to speed up the cool down time. The last stage (UC) and intermediate stage (IC) are connected by a passive graphite heat switch[7,8,9] which will turn OFF smoothly by itself as the instrument is cooling down. Indeed, the thermal conductivity of graphite is decreasing dramatically with temperature and we can consider that below 10 K, with a conductive ratio of 1/1000 comparing to room temperature conduction, the switch is OFF.

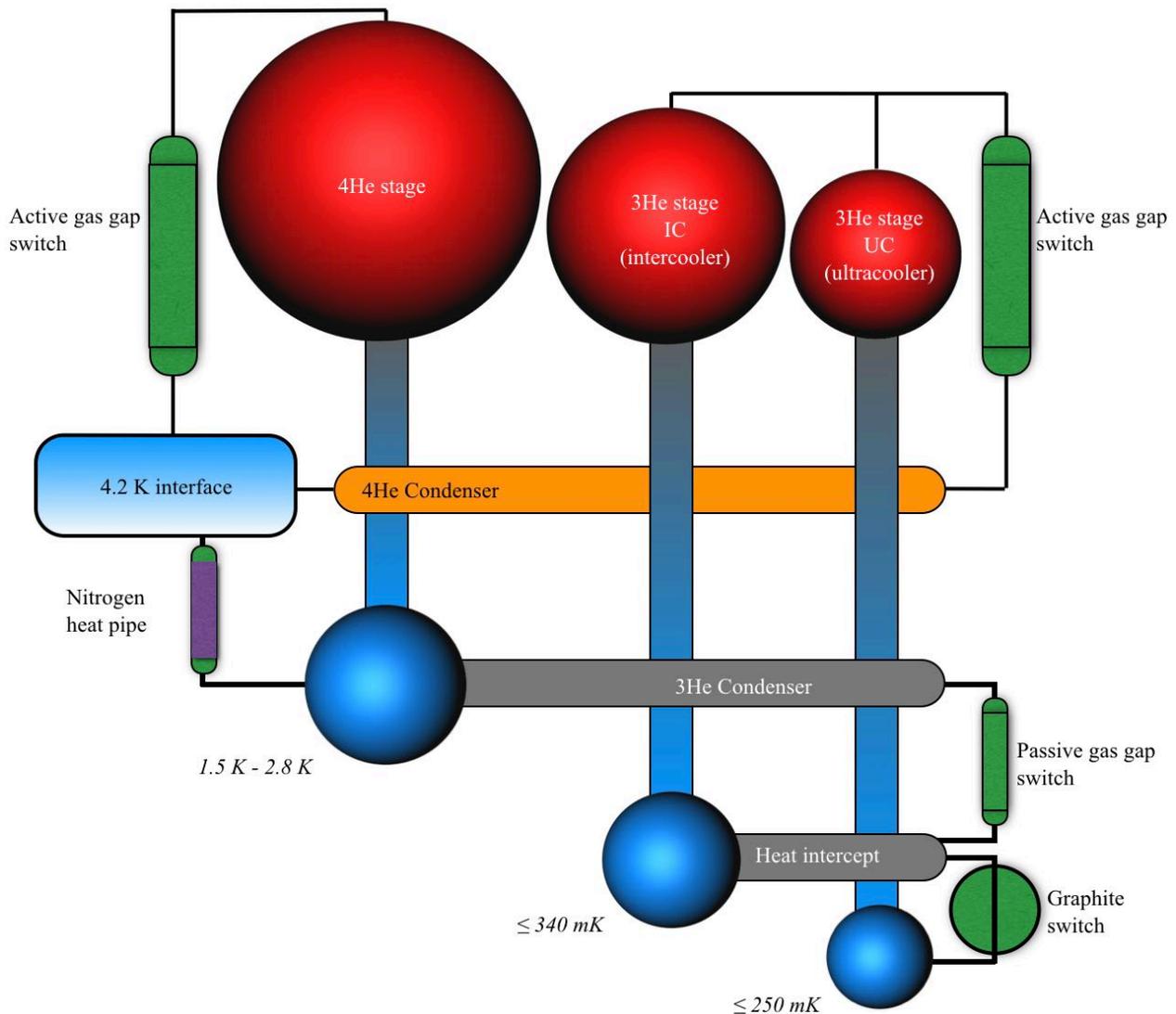

**Figure 3.** Schematic of the 3 stages cooler

The UC and $^4$He evaporators are connected by a passive gas gap heat switch, which will turn OFF in the 20 K range. The geometry of this switch has been chosen to lower the parasitic losses in the OFF mode The $^4$He evaporator is connected to the 4 K interface by a nitrogen heat pipe[10,11]. This will help cooling the $^4$He stage when the cryostat pulse tube is cooling in the 110 K - 63 K range. Above 110 K, nitrogen is not condensing and heat transfer is poor. Then below 63 K, at which temperature the nitrogen is frozen within the heat pipe and thus the heat transfer is not efficient anymore, the last part of the cooldown will rely on the gas in the tubes to the evaporators (including the above links). All these items are passive without any electrical connections.

Table 2 summarizes the main specifications as Fig. 4 shows 3D views of the cooler along with actual pictures of the assembled cooler. The cooler is a pressure device and all parts subject to internal pressure have been sized accordingly. The lowest burst pressure is expected to be around 20 MPa.

**Table 2** : cooler specifications (as build)

|  | First stage ($^4$He) | Second stage (IC) | Third stage (UC) |
|---|---|---|---|
| Helium charge | 68 STP dm$^3$ | 19.7 STP dm$^3$ | 4.3 STP dm$^3$ |
| Pressure at room temperature | 8 MPa | 7.34 MPa | 6.5 MPa |
| Overall dimensions | 202 x 210 x 219 mm | | |
| Overall mass | 6.7 kg | | |

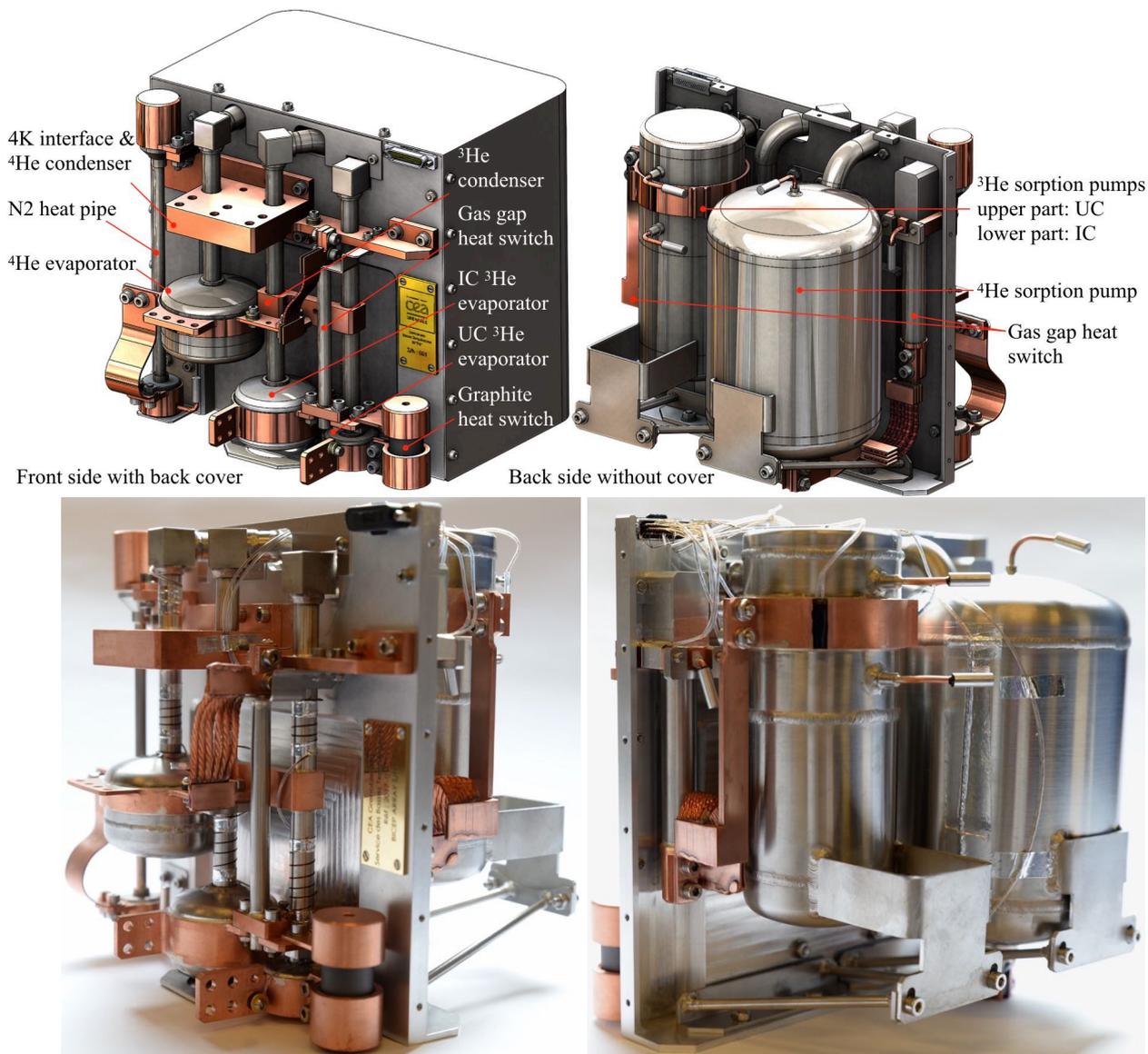

**Figure 4.** 3D views of the cooler and actual pictures

**EXPERIMENTAL PERFORMANCE**

**Cooldown**

The full cooler was integrated in the test cryostat (see Fig. 5). On the two cooldown curves depicted in Fig. 6, it can be seen that indeed the diode triggers at around 110 K leading to a much faster cooldown rate. Our test cryostat features a liquid nitrogen and a liquid helium tank. The first cooldown curve shown corresponds to the first phase of the cooldown, i.e. the nitrogen tank has been filled and the helium tank initially at 300 K is first carefully filled with liquid nitrogen to bring its temperature as close as possible to 80 K while making sure no liquid remains in the tank. For the subsequent phase, liquid helium is transferred and as soon as the temperature of the condenser drops below about 63 K it can be seen that the diode turns OFF.

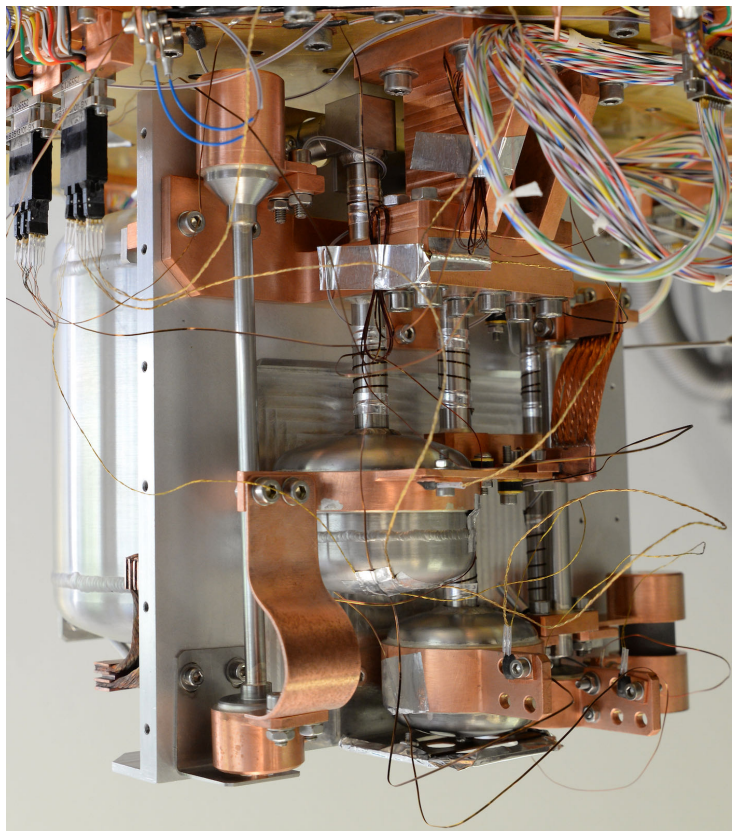

**Figure 5.** Cooler in the test cryostat

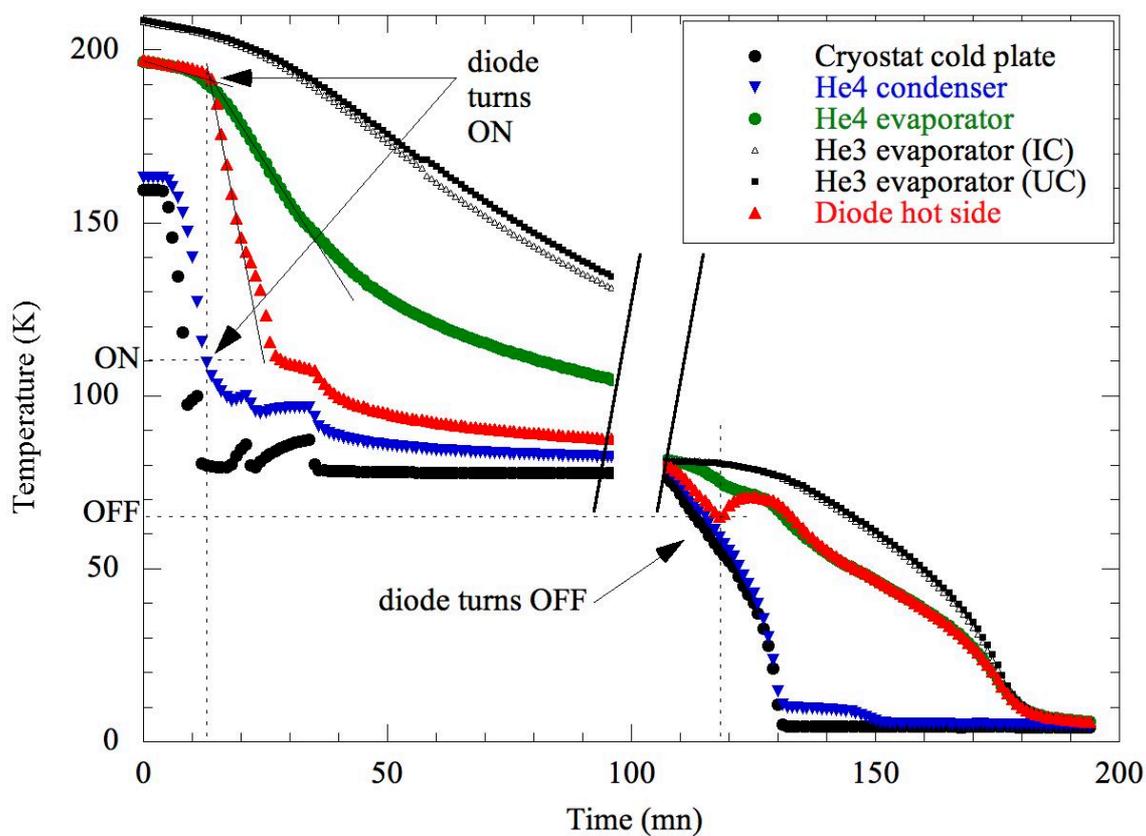

**Figure 6.** Cooldown curves

**Ultimate temperatures and cooling power curves**

The no-load temperatures as well as the temperature measured for the applied load specified in the requirements (see Table 1) are reported in Table 3. The cooling power curves under various conditions are reported in Fig. 7.

**Table 3**: Ultimate temperatures

| Stage | Ultimate temperature | Net heat lift / (requirement) |
|---|---|---|
| $^4$He stage 1$^{st}$ stage | 1010 mK | 230 μW @ 1130 mK / (≤ 2.8 K) |
| Intercooler (IC) $^3$He 2$^{nd}$ stage | 275 mK | 70 μW @ 317 mK / (≤ 340 mK) |
| Ultracooler (UC) $^3$He 3$^{rd}$ stage | 212 mK | 2 μW @ 227 mK / (≤ 250 mK)<br>15 μW @ 263 mK / (≈ 300 mK) |

**Hold time**

The autonomy of the cooler has been measured under various conditions, i.e. nominal applied loads as well as larger loads. From the calculated parasitic and applied load on each stage, one can calculate the energy available at the operating temperature and compare the result with the theoretical energy. The theoretical energy is based on the number of liters STP in each stage and assuming reasonable condensation efficiency and fraction lost during the cooldown process. This analysis shows that the numbers of Joules available at each stage is as expected. A full cycle under the nominal conditions is reported in Fig. 8.

The autonomy of the cooler exceeds the specified 48h because for these tests, no additional masses were connected to the various stages. The additional autonomy measured can be converted in Joules and compared to the expected enthalpies of the various materials considered (see Table 1). To calculate the additional Joules, we consider the applied loads on each stage as well as the parasitic loads, i.e. 376, 16 and 1 μW, respectively for the $^4$He, IC and UC stage. In addition, it should be noted that the hold time of the UC stage has been measured with 15 μW applied. In nominal operation the load is expected to be 2 μW, leading to 0.52 J (parasitic included) over 48 hours. Thus, the additional joules available for all three stages are 32, 4.2 and 2.7 respectively for the $^4$He, IC and UC stage, as the energies required to cool the various masses from 4.2 K are respectively 16.5, 0.4 and 1 Joules. It can be seen that once recycled, each stage holds enough energy to cooldown the large masses and to cover the 48 hours hold time with the specified loads.

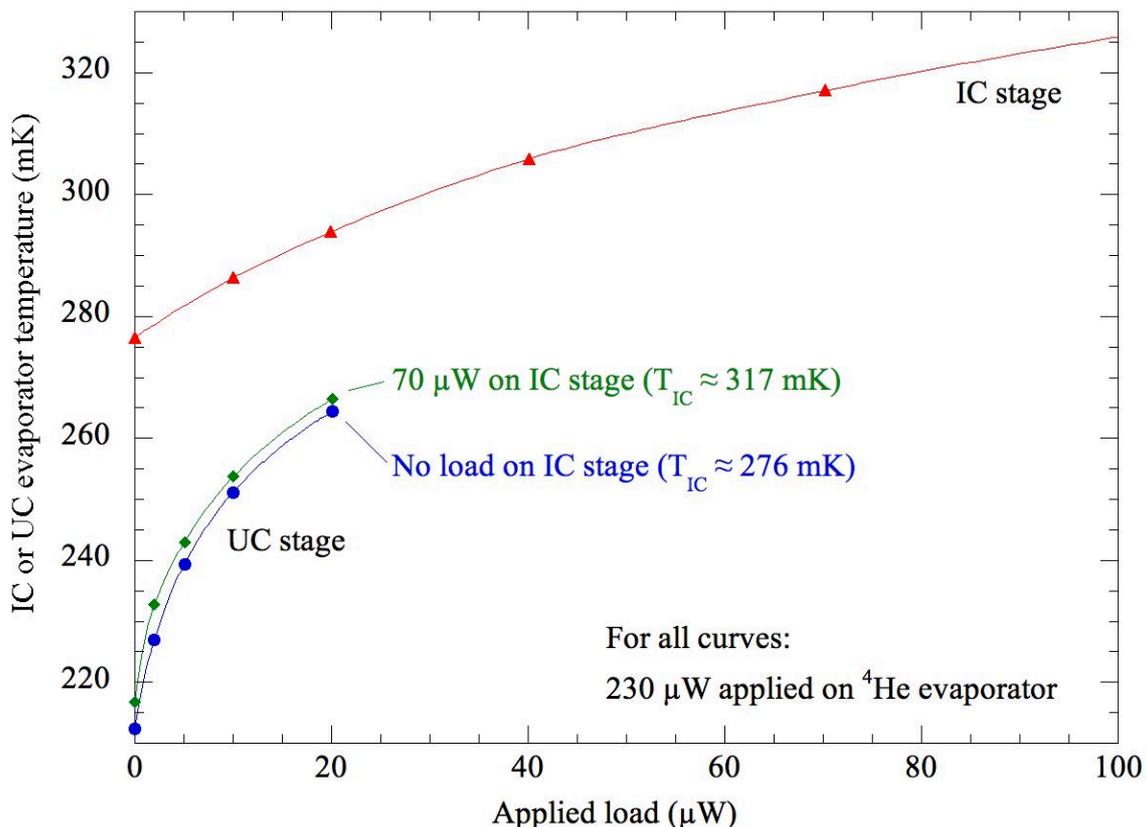

**Figure 7.** Cooling power of IC and UC stages

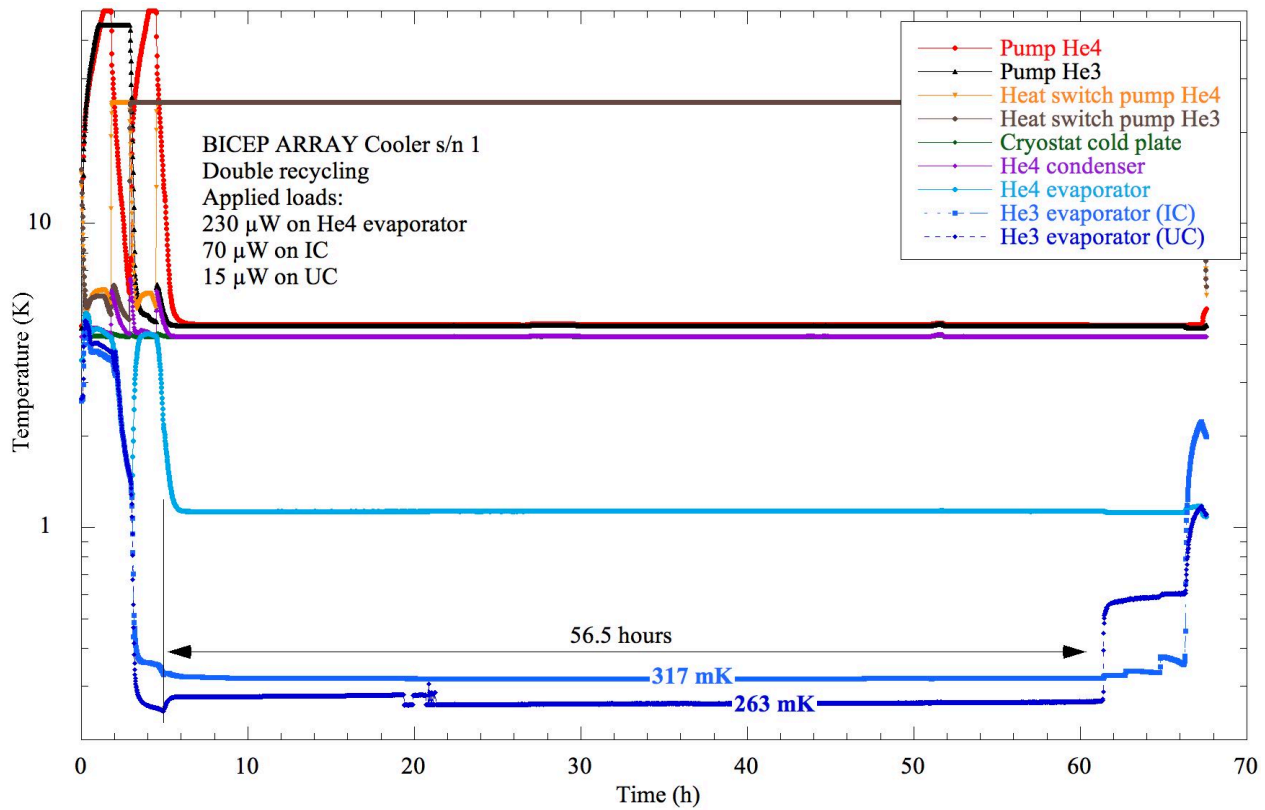

**Figure 8.** Full recycle with autonomy under the nominal applied loads

## CONCLUSION

A three stages sorption cooler has been sized, designed, assembled and tested for the BICEP Array project. This cooler will be used to cool to sub-kelvin temperatures several fairly massive items. The experimental results obtained show that the cooler is operating to specifications. This first cooler has been delivered to Caltech/JPL. 4 additional units are currently being assembled and will be tested in the coming months.

## ACKNOWLEDGMENT

We thank Florian Bancel and Laurent Clerc for their expert technical work on this project.

## REFERENCES

1. Kamionkowski, M. & Kovetz, E.D., "The Quest for B Modes from Inflationary Gravitational Waves" *Ann. Rev. Astron. Astrophys.* (2016), 54, 227.

2. Keck Array and BICEP2 Collaborations et al., "BICEP2 / Keck Array VI: Improved Constraints on Cosmology and Foregrounds When Adding 95 GHz Data from Keck Array" *Physical Review Letters* (2016), 116, 031302.

3. BICEP2 Collaboration et al., "BICEP2 III: Instrumental Systematics" *Astrophysical Journal* (2015), 814, 110.

4. Duband, L., Lange, A. and Ravex, A. "A miniature adsorption 3He refrigerator" *Proceedings of the 4th European Symposium on Space Environmental and Control Systems*, October 21-24 1991, Florence, Italy (ESA SP-324, December 1991) p. 407-409

5. Duband, L. "Double stage helium sorption coolers" *Cryocoolers 11, Proceedings of the 11$^{th}$ International Cryocooler Conference,* June 20-22 2000, Keystone CO USA

6. Ercolani, E et al., "Cryogenic system for the ArTeMiS large sub millimeter camera" *Proc. SPIE 9153, Millimeter, Submillimeter, and Far-Infrared Detectors and Instrumentation for Astronomy VII*, 915324 (July 2014)

7. M.C. Runyan, et al., "Thermal conductivity of thermally-isolating polymeric and composite structural support materials between 0.3 and 4 K" *Cryogenics*, v.48 (2008), pp. 448–454
8. A.L. Woodcraft, et al, "A replacement for AGOT graphite?," *Physica B*, 329-333 (2003), pp. 1662-1663
9. Prouvé & al., "A 1K 4He Close Cycle Loop Precooled Using a PT415 Pulse Tube for the BLISS Test Bed Cryostat" *Cryocooler 17*, International Cryocooler Conference, Boulder CO, (2012), 487-49
10. Bewilogua & al., "Application of the thermosiphon for precooling of apparatus" *Cryogenics* 1966, 34-35
11. Prouvé & al., "Pulse-tube dilution refrigeration below 10 mK" *J. of Low Temp. Phys*., 2007, 148 (5-6), pp.909, https://hal.archives-ouvertes.fr/hal-00921527